\pdfoutput=1
\documentclass[aps,prb,twocolumn,showpacs,amssymb]{revtex4}

\usepackage{graphicx}
\usepackage{dcolumn}

\usepackage{bm}
\def\be{\begin{equation}}
\def\en{\end{equation}}

\begin{document}

\title{Structure formation of surfactant membranes under shear flow }


\author{Hayato Shiba}\email[]{shiba@issp.u-tokyo.ac.jp}
\author{Hiroshi Noguchi}\email[]{noguchi@issp.u-tokyo.ac.jp}
\affiliation{Institute for Solid State Physics, University of Tokyo, 
Chiba 277-8581, Japan
}
\author{Gerhard Gompper}
\affiliation{Theoretical Soft Matter and Biophysics,
Institute of Complex Systems and Institute for Advanced Simulation,
Forschungszentrum J\"ulich, D-52425 J\"ulich, Germany}


\date{\today}

\begin{abstract}
Shear-flow-induced structure formation in surfactant-water mixtures 
is investigated numerically using a meshless-membrane model in
combination with a particle-based hydrodynamics simulation approach
for the solvent.
At low shear rates, 
uni-lamellar vesicles and planar lamellae structures are formed 
at small and large membrane volume fractions, respectively. 
At high shear rates,
lamellar states exhibit an undulation instability,
leading to rolled or cylindrical membrane shapes oriented in the 
flow direction.
The spatial symmetry and structure factor of this rolled state
agree with 
those of intermediate states during lamellar-to-onion transition 
measured by time-resolved scatting experiments.
Structural evolution in time exhibits 
a moderate dependence on the initial condition. 
\end{abstract} 


\maketitle  


\section{Introduction}
\label{sec:intro}

Surfactant molecules in water self-assemble into various structures
such as micelles and bilayer membranes, which display a rich variety 
of rheological properties under flow. \cite{lars99b}
Even if a basic structure remains to be a bilayer membrane,
its mesoscale structure can assume several different states, such
as fluid $L_\alpha$ or ripple $P_\beta$ phase.
Under shear flow, lamellae can be oriented parallel or perpendicular 
to the shear-gradient direction.
Diat and Roux first discovered 20 years ago 
closely-packed multi-lamellar vesicle (MLV) structures, 
so-called the onion phase, 
in nonionic surfactant-water mixtures under shear flow. \cite{93Diat,93RouxEPL,95Diat} 
In the last two decades, this onion structure
has been studied experimentally using 
light, \cite{93Diat,95Diat,02Panizza,09Richtering,10Suganuma,12Fujii} 
neutron, \cite{93Diat,95Diat,03Nettesheim}  
and X-ray scattering, \cite{10Suganuma,11Ito,12Fujii}
and also by freeze-fracture electron microscopy, \cite{02Panizza} 
and the rheo-NMR method. \cite{08Medronho,09Medronho,10Medronho}
Its rheology has been of large 
interest. \cite{93RouxEPL,01Richtering,03Nettesheim,07Miyazawa,09Richtering,12Fujii} 
Typically, a critical shear rate $\dot{\gamma}_c$ separates
the lamellae and onion phases, where the latter phase consists of 
mono-disperse onions containing hundreds of lamellar layers.
The onion radius $R(\dot{\gamma})$ is reversible and 
can be described by a unique decreasing function 
of the shear rate $\dot{\gamma}$. \cite{05Richtering}

Time-resolving small-angle neutron scattering experiments have revealed 
that a two-dimensional intermediate structure is formed during the 
lamellar-to-onion transition with increasing shear rate. \cite{03Nettesheim}
A cylindrical or wavy lamellar structure was speculated to be the 
transient intermediate structure, but could not be distinguished from 
the scattering pattern alone.
Recent small-angle X-ray scattering experiments with increasing temperature 
at constant shear rate also indicate a similar pattern
around the lamellar-to-onion transition. \cite{11Ito}
Thus, there are some experimental evidences of a 
transient state, but its structure is still under debate.  
An alternative experimental approach to gain insight into the 
structural changes is to characterize defects observed in 
the lamellar state for moderate shear rates, both in 
surfactant membranes \cite{10Medronho,11Medronho}
and in 
thermotropic liquid crystals. \cite{10Fujii,11Fujii} 
It is also worth mentioning that stable cylindrical structures on 
a ten-micrometer length scale are observed when strong shear flow 
is applied in the lamellar-sponge coexistence state. \cite{07Miyazawa}

Several theoretical attempts have been made to tackle 
this complex problem of structural evolution under shear flow,
which consider either instability of the lamellar phase due to
undulations \cite{ZG,02Olmsted} or the break-up of 
droplets. \cite{96vdLinden,93vdLinden}
Recently, a ``dynamical'' free energy of MLV under shear flow 
has been proposed, \cite{LuPRL} which takes
into account the slow modes induced by the solvent between the membranes
together with their bending and stretching forces. 
The scaling relations for the MLV size and the terminal shear rate are 
predicted in agreement with the experiments of 
Refs.~\onlinecite{93Diat,93RouxEPL}.
In these theories, while the hydrodynamic effects 
of the solvent are taken into account, 
the analysis is performed for geometrically simple structures, asp
spherical onions or planar lamellae.
Thus, the kinetic process of the transformation from the lamellar to 
the onion phase could not be investigated theoretically so far.
In this paper, we study the detailed structural evolution
of surfactant membranes under simple shear flow
using large-scale particle simulations.

A few simulations have been performed 
for the formation of lamellar phases in shear flow,
while onion formation has never 
been addressed so far. 
Oriented lamellae have been obtained in simulations of 
a coarse-grained molecular model for lipids, \cite{02Guo,04Soddemann,07Guo}
while defect dynamics has been investigated in simulations of a 
phase-field model of a smectic-A system. \cite{12Coveney}
Onion and intermediate states
have large-scale structures of the order of micrometers,
which are beyond typical length scales accessible to molecular dynamics 
(MD) simulations of coarse-grained surfactant molecules.
In our study, we employ a meshless-membrane model, 
\cite{NogRev,Nog06PRE,Nog06JCP,11Shiba}
where a membrane particle represents not a surfactant molecule but rather
a patch of bilayer membrane -- in order to capture the membrane dynamics
on a micrometer length scale.
This model is well suitable to study membrane dynamics accompanied 
by topological changes.
Alternatively, membranes can be modeled as triangulated surfaces, 
\cite{NogRev,gg:gomp04c,fedo13} 
which require however discrete bond reconnections to describe 
topological changes. \cite{gg:gomp98c,gg:gomp04c,gg:gomp12g}
After the first meshless-membrane model was proposed in 1991, \cite{Drouffe}
several meshless-membrane models have been developed. 
\cite{Nog06PRE,Nog06JCP,11Shiba,popo08,09Kohyama,2010Yuan1,10PLGeissler}
In contrast to other meshless-membrane approaches, our models 
\cite{Nog06PRE,Nog06JCP,11Shiba} are 
capable of separately controlling the membrane bending rigidity 
$\kappa$ and the line tension $\Gamma$ of membrane edges.
Previously, \cite{NogRev,Nog06JCP} we have combined our 
meshless-membrane model with multi-particle collision (MPC) dynamics, 
\cite{kapr08,gg:gomp09a} a particle-based hydrodynamic simulation technique. 
With MPC, the hydrodynamic interactions are properly taken into account,
but due to the frictional coupling of membrane and solvent, solvent 
particles can penetrate through the membrane.
We here extend the meshless-membrane model 
into explicit solvent simulation model, in which the fluid particles 
interact with each other and with the membrane via short-ranged 
repulsive potentials, so that the solvent can hardly penetrate the membrane,
and simulate it with dissipative particle dynamics (DPD), 
another hydrodynamics simulation technique.

The simulation model and methods are introduced in Sec.~\ref{sec:model}.
Then basic membrane properties, including bending rigidity and  
line tension, are described in Sec.~\ref{sec:prop}.
In Sec.~\ref{sec:results}, structure formation in surfactant-water
mixtures under shear flow is investigated for variety of shear rates 
$\dot{\gamma}$ and membrane volume fractions $\varphi$. 
At high $\dot{\gamma}$ and high $\varphi$, a novel structure of rolled-up
membranes is found, which are oriented in the flow direction. 
A summary and some perspectives are given in Sec.~\ref{sec:sum}.

\section{Model and Methods}
\label{sec:model}

\subsection{Coarse-Grained Model and Interaction Potentials}

To simulate the structure formation in surfactant-membranes systems,  
we employ a meshless-membrane model with explicit solvent.
In this model, two types of particles $\mathcal{A}$ 
and $\mathcal{B}$ are employed, which denote membrane 
and solvent particles, respectively. 
The number of these particles 
is $N_\mathcal{A}$ and $N_\mathcal{B}$, which defines the 
particle density $\phi=(N_\mathcal{A}+N_\mathcal{B})/V$ -- where $V$ is
the volume of the simulation box -- and the membrane
volume fraction $\varphi=N_\mathcal{A}/(N_\mathcal{A}+N_\mathcal{B})$.
The particles interact via a potential $U$, which consists of
repulsive, attractive, and curvature interactions, 
$U_{\rm rep},\ U_{\rm att},$ and $U_\alpha$, respectively,
\begin{equation}
\frac{U}{k_{\rm B}T} = \sum_{i<j}  U_{\rm rep} + 
\sum_{i\in \mathcal{A}} \epsilon U_{\rm att} +  k_\alpha U_\alpha.
\end{equation}
Here, the former sum for repulsive interactions 
is taken over all pairs of particles, the latter only over the membrane 
particles. 

All neighbor particle pairs interact via the short-ranged repulsive 
potential
\begin{equation}
U_{\rm rep} = \left\{ 
\begin{array}{ll}
\epsilon_c \left( \frac{\sigma}{r_{ij}} \right)^{12} - B & (r_{ij}<r_c^{\rm rep}) \\
0 & (r_{ij} \ge r_c^{\rm rep})
\end{array}
\right.
\end{equation}
where $r_{ij}$ is the distance between particles $i$ and $j$. 
This potential is cut off at a distance $r_c^{\rm rep} = 3.2\sigma$.
The length $\sigma$, representing the particle  diameter, 
is employed as the length unit. 
The constant $B$ is chosen such as to ensure the continuity of the 
potential at $r=r_c^{\rm rep}$. The (dimensionless) potential strength 
is set to $\epsilon_c=4$.

To favor the assembly of membrane particles into smoothly curved 
sheets in three-dimensional (3D) space, 
membrane particles interact via the additional potentials $U_{\rm att}$
and $U_\alpha$,
which have been introduced in the implicit-solvent version of the 
model previously. \cite{Nog06PRE,Nog06JCP,11Shiba}
With these potentials, the membranes particles self-assemble into a 
single-layer sheet, which is a model representation of a bilayer 
membrane.  Here, the attractive interaction is given by 
\begin{equation}
U_{\rm att} =  0.25 \ln \{ 1+ \exp [-4 (\rho_i -\rho^* ) ] \} - C,  
\end{equation} 
which is a function of the local density $\rho_i$ of the membrane 
particles defined by 
\begin{equation}
\rho_i = \sum_{i\in \mathcal{A}} f_{\rm cut} (r_{ij}/ \sigma ). \label{eq:rho}
\end{equation}
$C=0.25 \ln [1+\exp (4\rho^*)]$ is chosen such that $U_{\rm att}(0) = 0$.
The cutoff function $f_{\rm cut}$ in Eq.~(\ref{eq:rho}) 
is a $C^\infty$ function represented as 
\begin{equation}
f_{\rm cut}(s) = \left\{ 
\begin{array}{ll}
\exp \left[ a (1+ \frac{1}{ (|s| /s_{\rm cut} )^n -1} ) \right] & (s<s_{\rm cut}) \\
0 & (s\ge s_{\rm cut})
\end{array}
\right.
\end{equation}
where  $n=12$, 
$a=\ln(2) \{(s_{\rm {cut}}/s_{\rm {half}})^n-1\}$ with
$s_{\rm cut}=2.1$ and $s_{\rm {half}}=1.8$ 
are used. 
We set $\rho^* = 6$ to study a fluid membrane in an explicit solvent. 

The curvature potential 
\begin{equation}
U_\alpha = \sum_{i\in\mathcal{A}} \alpha_{\rm pl} (\bm{r}_i) 
\end{equation}
is introduced to incorporate the membrane bending rigidity.
Here, the aplanarity $\alpha_{\rm pl}$ provides a 
measure for the degree of deviation of membrane 
particle alignment from a planar reference state. It is defined by
\begin{equation}
\alpha_{\rm pl} = \frac{9D_{\rm w}}{T_{\rm w}M_{\rm w}}
= \frac{9\lambda_1 \lambda_2 \lambda_3}{(\lambda_1+\lambda_2+\lambda_3)
(\lambda_1 \lambda_2 + \lambda_2\lambda_3 +\lambda_3 \lambda_1)}, 
\end{equation}
where $\lambda_1 \le \lambda_2 \le \lambda_3$ are 
the three eigenvalues of the local gyration tensor 
$a_{\alpha\beta}$  of the membrane near particle $i$, which is defined as 
$a_{\alpha\beta} = \sum_{j\in\mathcal{A}} 
(\alpha_j -\alpha_G)(\beta_j -\beta_G) w_{\rm cv} (r_{ij})$,
with $\alpha,\beta \in \{ x,y,z\}$. 
Here, $\bm{r}_G = \sum_{j\in\mathcal{A}} \bm{r}_j w_{\rm cv}(r_{ij}) / 
\sum_{j\in\mathcal{A}} w_{\rm cv}(r_{ij})$ 
is the locally weighted center of mass, and 
$w_{\rm cv} (r_{ij})$ is a truncated Gaussian function 
\begin{equation}
w_{\rm cv}(r_{ij} ) = \left\{ \begin{array}{ll}
\exp \left( \frac{  (r_{ij}/r_{\rm ga})^2 }{ (r_{ij}/r_{\rm cc})^n -1} \right) & (r_{ij}<r_{\rm cc} ) \\
0 & (r_{ij} \ge r_{\rm cc} ) 
\end{array} \right.  \label{eq:wcv}
\end{equation}
which is smoothly cut off at $r_{ij} = r_{\rm cc}$. 
The constants are set as $n=12,\ r_{\rm ga} = 1.5\sigma$, and 
$r_{\rm cc} = 3.2\sigma$. 

In all our simulations, the volume is chosen such that 
the number density $\phi$ of the particles is constant,
\begin{equation}
\phi = N/V =  0.64\sigma^{-3},\ N=N_{\mathcal{A}} + N_{\mathcal{B}}.
\end{equation}
For higher solvent densities, the system is closer to the melting point,
and strong attractive interaction between membranes sheets have been
found (see Appendix A for details). 
The density $\phi = 0.64\sigma^{-3}$ is chosen in order to avoid such an attraction.

\subsection{Thermostats}

We simulated the membranes in the NVT ensemble under shear flow. 
To keep the temperature constant, we employ  
the dissipative particle dynamics (DPD) 
thermostat, \cite{hoog92,95Espanol,groo97,07Nog,07Nog2}
in which friction and noise forces are applied to the relative
velocities of pairs of neighboring particles. 
Thus, linear and angular momentum is conserved, which implies 
that the system shows hydrodynamic behavior on sufficiently 
large length and time scales. 
The equation of motion for the $i$th particle is given by 
\begin{equation}
m\frac{d\bm{v}_i}{dt} = -\frac{\partial U}{\partial\bm{r}_i}
+ \sum_{j\neq i} \{ -w_{ij} (\bm{v}_i-\bm{v}_j)
\cdot  \hat{\bm{r}}_{ij} 
+\sqrt{w_{ij}} \xi_{ij}(t) \} \hat{\bm{r}}_{ij} ,
\end{equation}
where $\hat{\bm{r}}_{ij} = \bm{r}_{ij} /r_{ij}$.
Here, the weight function $w_{ij}$ is 
$w_{ij}(r_{ij})  = \gamma \theta ( A \sigma - r_{ij})$
with $A=2.7$, where $\theta(r)$ is the Heaviside step function. 
This type of weight has also been 
used in the Lowe-Andersen thermostat. \cite{lowe99} 
The scheme is discretized with Shardlow-S1
splitting method, \cite{Shardlow} whose time step 
$\Delta t^b = 0.2 t_0$ is different from that for 
contributions from the molecular interactions, where 
$\Delta t=0.005t_0$ ($t_0=m/\gamma$ is simulation time unit).
In the following, 
we use $\gamma = \sqrt{mk_{\rm B}T}/\sigma$,
where $k_{\rm B}T$ is the thermal energy unit.  
Although the DPD thermostat is usually employed for  
soft interaction potentials, \cite{09MarrinkRev,SmitRev,04Laradji}
it can also be employed for systems with steeper potentials, 
such as the Weeks-Chandler-Andersen potential. \cite{04Soddemann,02Guo}

Shear flow with velocity $v_x=\dot{\gamma}z$ in the $x$-direction and
gradient in the $z$-direction is imposed by Lees-Edwards boundary condition. 
The code is optimized for use on a parallel computer architecture 
by domain decomposition (see Appendix B).

\section{Model Properties}
\label{sec:prop}

The meshless-membrane model with explicit solvent 
is expected to exhibit very similar equilibrium properties as the 
original implicit-solvent version. \cite{Nog06PRE,Nog06JCP,11Shiba}
We now confirm this dependence of surface tension, line tension, and 
bending rigidity on the control parameters for the explicit-solvent 
model.

\begin{figure}
\includegraphics[width=0.85\linewidth, bb=0 0 360 252]{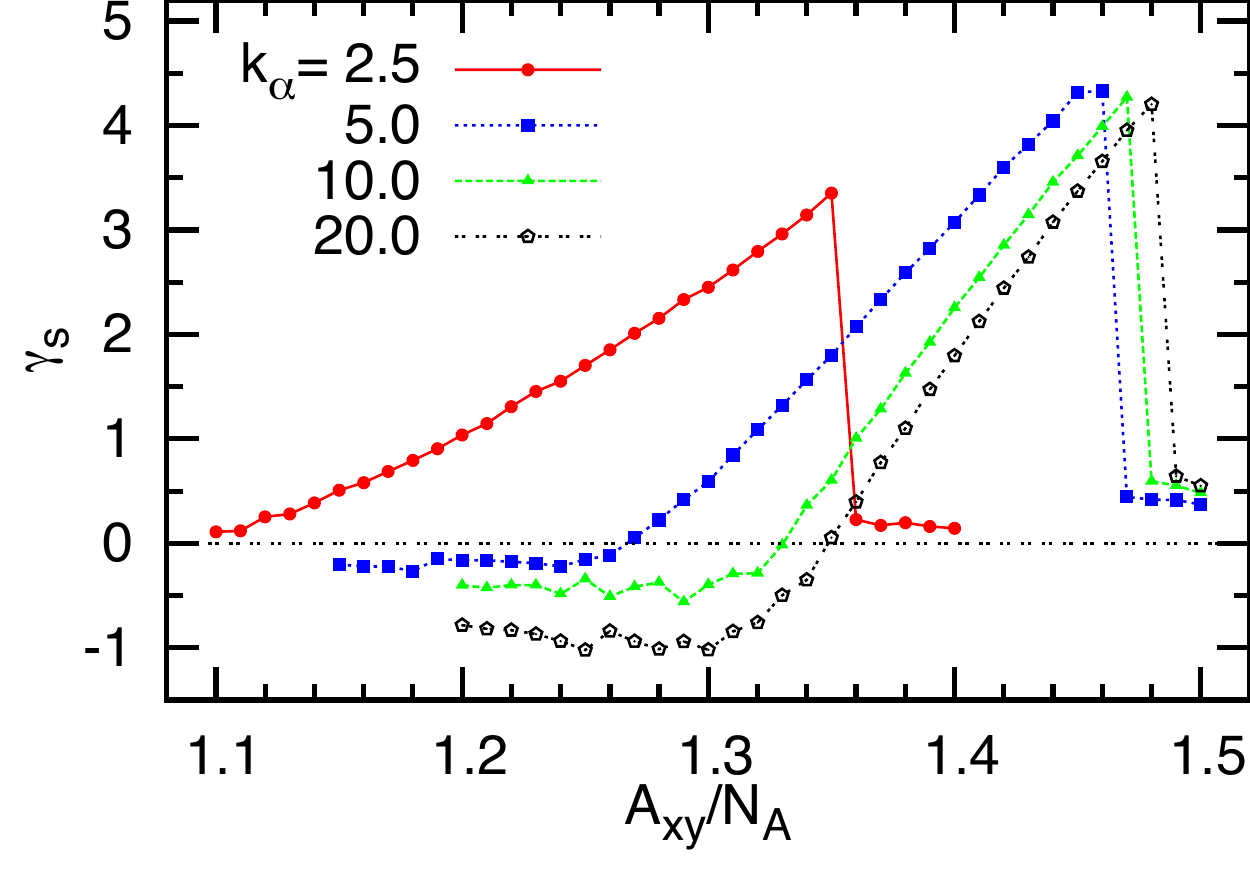}
\caption{\label{fig:SurfT}
Area dependence of surface tension $\gamma_{\rm s}$ 
for the explicit solvent meshless membrane  model is plotted for 
$k_\alpha = 2.5, 5, 10,$ and 20
with $\epsilon=4$. 
}
\end{figure}

First, a planar fluctuating membrane is simulated with the 
particle numbers $N_\mathcal{A}=1600$ (and $N=48\ 000$
or $64\ 000$). 
For various membrane projected areas $A_{xy}=L_xL_y$,
the surface tension 
\begin{equation}
\gamma_{\rm s} = \langle P_{zz} - (P_{xx}+P_{yy})/2 \rangle L_z 
\label{eq:sftens} 
\end{equation}
is investigated. Here, $P_{\mu\nu}$ is the pressure tensor 
given by 
\begin{equation}
P_{\mu\nu } = \sum_{i=1}^N  \left( mv_i^\mu v_i^\nu - \mu_i 
\frac{\partial U}{\partial \nu_i}\right) \Big/ V,
\end{equation}
where $\{\mu, \nu\} \in \{x,y,z\}$, $\bm{v}_i = (v_i^x, v_i^y, v_i^z)$, 
and the sum is taken over all the particles including
the solvent component. 
In calculating $P_{\mu\nu}$, the periodic image 
$\mu_i + nL_\mu$ nearest to the other interacting 
particles is employed, when the potential 
interaction crosses one of the periodic boundaries. 

Figure \ref{fig:SurfT} shows the dependence of
$\gamma_{\rm s}$ on the projected membrane area $A_{xy}$ 
for $k_\alpha = 2.5, 5, 10$, and 20, with $\epsilon = 4$. 
For $\gamma_{\rm s} \simeq 0$, the intrinsic area $A$ is larger 
than $A_{xy}$ due to the membrane undulations; buckling of the 
membrane occurs for $A < A_{xy}$ (the flat region at 
$\gamma_{\rm s} < 0$ in Fig.~\ref{fig:SurfT}). \cite{nogu11a}
For tension-less membranes with $N_\mathcal{A}=1600$, the 
projected area $A^0_{xy}$ is  given by 
$A_{xy}^0 = a^0_{xy} N_\mathcal{A}$ with 
$a_{xy}^0 = 1.1,\ 1.27,\ 1.33,\ 1.35$
for $k_\alpha = 2.5,\ 5,\ 10,\ 20$, respectively. 
 The projected membrane area increases  
with increasing $k_\alpha$,
because both membrane bending fluctuations and protrusions 
are suppressed at larger $k_\alpha$.

\begin{figure}
\includegraphics[width=0.8\linewidth, bb=0 0 360 330]{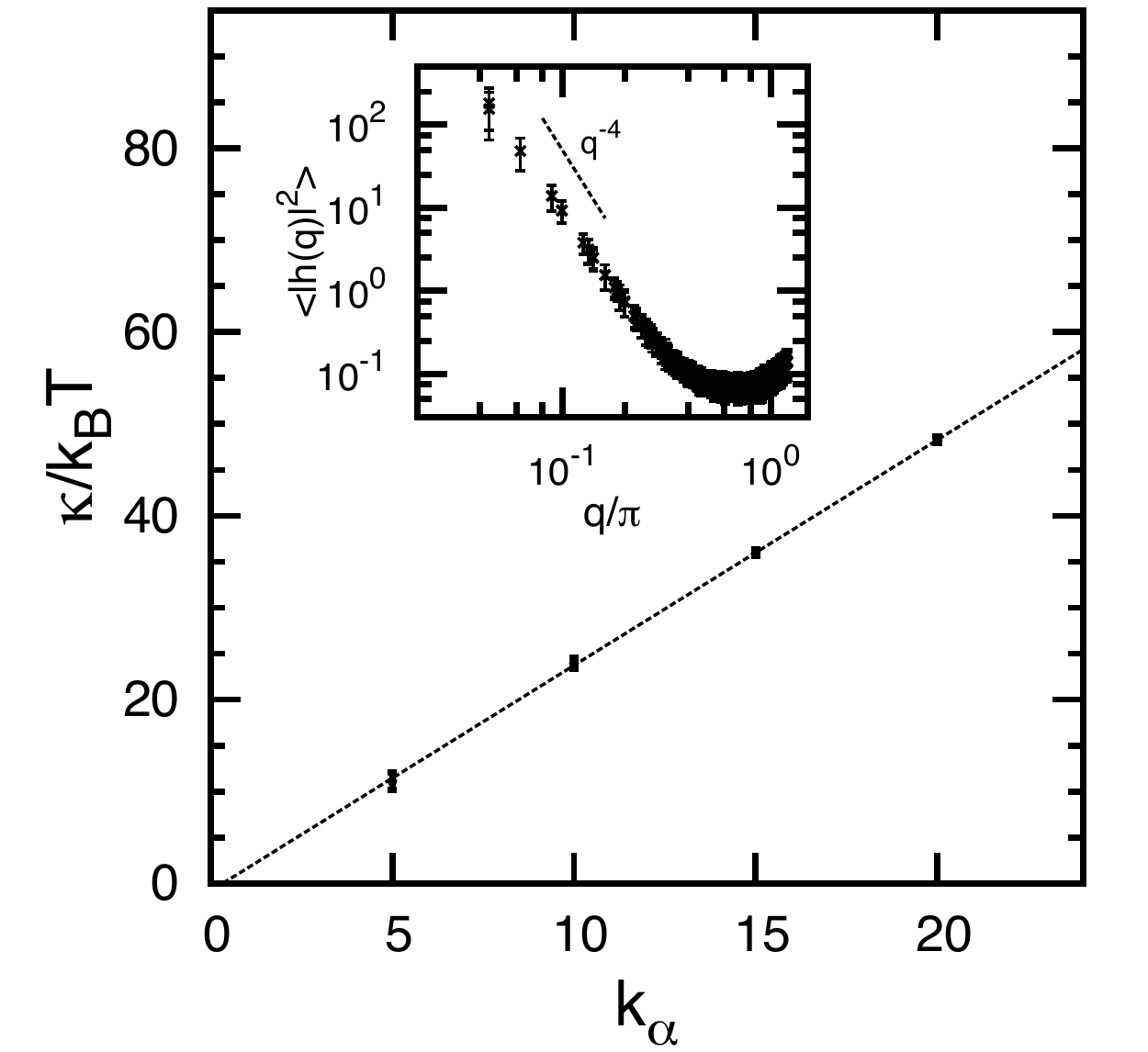}
\caption{\label{fig:kalpha}
Dependence of the bending rigidity $\kappa$ 
on $k_\alpha$ is plotted for $\epsilon = 4$, estimated
with the use of a tension-less planar membrane at $N_{\mathcal{A}}
= 1600$ through Eq.~(\ref{eq:Helfrich}). 
As shown in the inset, the height spectrum of 
the membrane exhibits a $q^{-4}$ spectrum.
}
\end{figure}

\begin{figure}
\includegraphics[width=0.75\linewidth, bb=0 0 360 252]{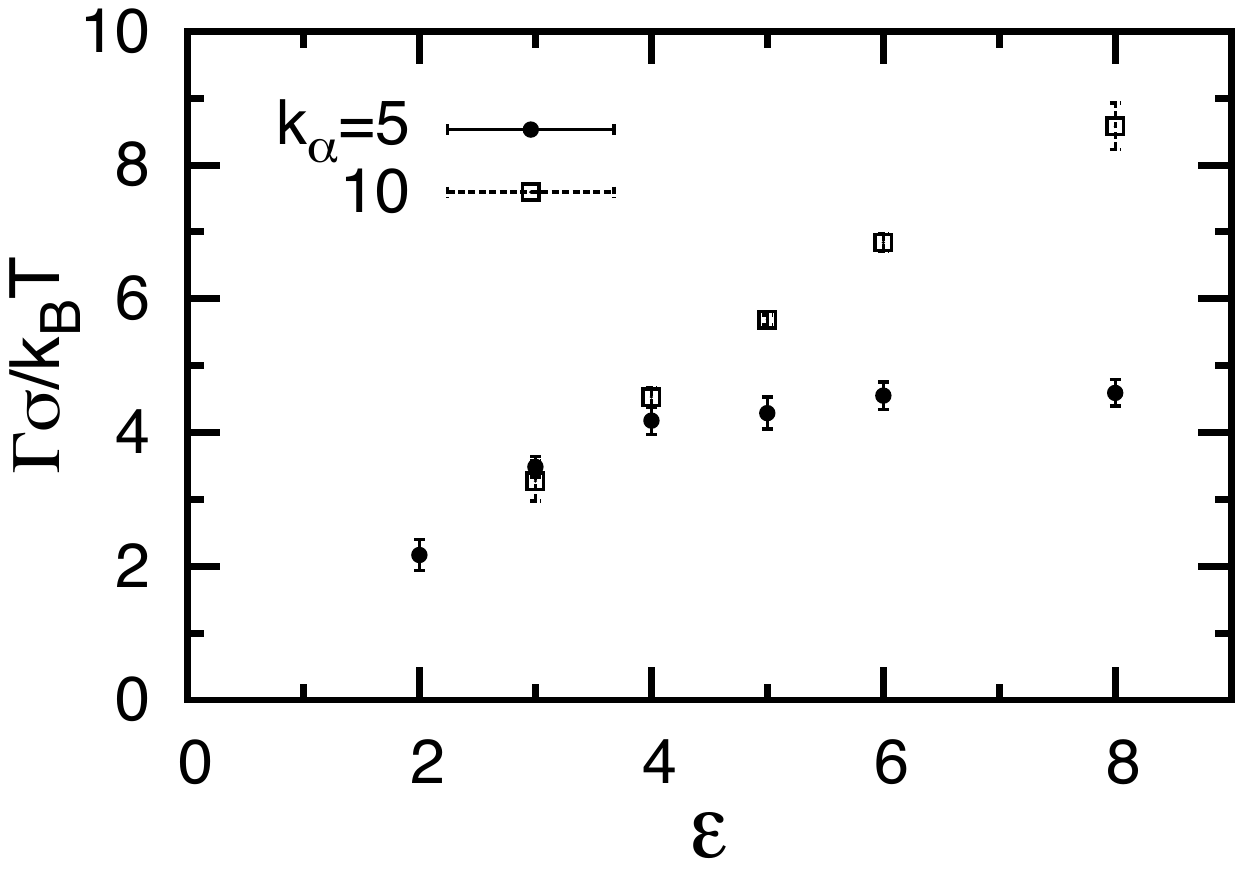}
\caption{\label{fig:LineT}
$\epsilon$ dependence of the line tension $\Gamma$ 
of explicit-solvent meshless-membrane  model, calculated for
$k_\alpha = 5$ and $10$ with the use of Eq.~(\ref{eq:linet}).
}
\end{figure}

Figure \ref{fig:kalpha} shows the bending rigidity $\kappa$ 
as a function of $k_\alpha$.
Here, the bending rigidity is estimated from
the height spectrum \cite{Helfrich}  
\begin{equation}
\langle |h(q)|^2\rangle = \frac{k_{\rm B}T}{\gamma_{\rm s} q^2 + \kappa q^4}. 
\label{eq:Helfrich}
\end{equation}
of the tension-less membrane ($\gamma_{\rm s}=0$).
In calculating $\langle |h(q)|^2\rangle$, the raw positional data 
of the membrane particles ($\bm{r}_i,\ i\in\mathcal{A}$) 
is employed. \cite{Nog06PRE,11Shiba}
Because of the slow dynamics of long-wavelength height fluctuations, 
it becomes more time-consuming to obtain precise data than 
for the implicit-solvent model. \cite{Nog06PRE,11Shiba}
Therefore, $\langle |h(q)|^2\rangle$
is measured using 16 independent runs for each $k_\alpha$,
and the averaged spectrum is then fitted to Eq.~(\ref{eq:Helfrich}) 
in the range $q < 1.2\sigma^{-1}$. 
As in the implicit model, the bending rigidity is found to be 
proportional to $k_\alpha$, which demonstrates the 
controllability of $\kappa$ in our model. 

In Fig.~\ref{fig:LineT}, the $\epsilon$ dependence
of the line tension $\Gamma$ is displayed for $k_\alpha = 5$ and 10  
for a membrane strip with two edges with lengths
equal to $L_x$ with $N_{\mathcal A} = 1600$.
Here, $\Gamma$ is determined via the relation \cite{11Shiba}
\begin{equation}
\Gamma = \langle (P_{yy} + P_{zz})/2 - P_{xx}\rangle  L_yL_z/2.    
\label{eq:linet}
\end{equation} 
At a small $\epsilon$, $\Gamma$ is proportional to $\epsilon$
and almost independent of $k_\alpha$, similarly to
the implicit-solvent meshless membrane  model \cite{Nog06PRE}. 
While $\Gamma$ increases linearly up to $\epsilon = 8$ for $k_\alpha = 10$,
it levels off and saturates at $\epsilon =4$ for $k_\alpha=5$. 
When $\epsilon$ exceeds the value where $\Gamma$ saturates
--- a value which becomes larger for larger $k_\alpha$ ---
the membrane particles prefer to reside at the edges, because the 
curvature force is not strong enough to avoid
aggregation due to the stronger attractive forces. 

For our simulations of membrane ensembles in shear flow, we choose 
the model parameters $k_\alpha = 5$ and $\epsilon = 4$,
 where the membrane has    
bending rigidity $\kappa/k_BT = 11\pm 1$. 
With the estimated values for the 
line tension $\Gamma\sigma /k_{\rm B}T = 4.2 \pm 0.2$, 
the relaxation time scale 
of structural transitions can be characterized by 
\begin{equation}
\tau = \frac{\eta R_c^3}{\kappa}, 
\end{equation}
where $\eta$ is the solvent viscosity, and $R_c$ is the critical 
radius of a flat disk. Assuming a flat disk with radius $R$ 
and a vesicle with the same membrane 
area, we obtain the corresponding elastic 
free energies $\mathcal{F}_d = 2\pi R\Gamma$ and 
$\mathcal{F}_s = 8\pi (\kappa + \bar{\kappa} /2)
\simeq 4\pi \kappa$; thus, a disk exhibits transition
to closed vesicle shape if the radius 
is around 
\begin{equation}
R_c = 2\kappa /\Gamma \sim (5.3 \pm 0.5 )\sigma . 
\label{eq:r_c}
\end{equation}
With the solvent viscosity 
$\eta = (2.1\pm 0.1)\times m(\sigma t_0)^{-1}$
obtained from a solvent-only simulation, 
an estimation of the relaxation time yields
$\tau = 28.4t_0$. 
In the following, the time will be  measured in unit of $\tau$. 
Since the membrane thickness, typically around 5nm for non-ionic
surfactants like polyethylene-glycol-ethers C$_n$E$_m$, 
corresponds to the size $\sigma$ of the membrane particles
in our simulation, $\tau$ is equivalent to about $0.36\mu$s
(with the viscosity of water at 300K,
$\eta_{\rm w} \simeq 0.8 {\rm mPa}\cdot{\rm s}$).

\section{Structure Formation with and without Shear Flow}
\label{sec:results}

\begin{figure*}
\includegraphics[width=0.80\linewidth, bb = 0 0 367 257]{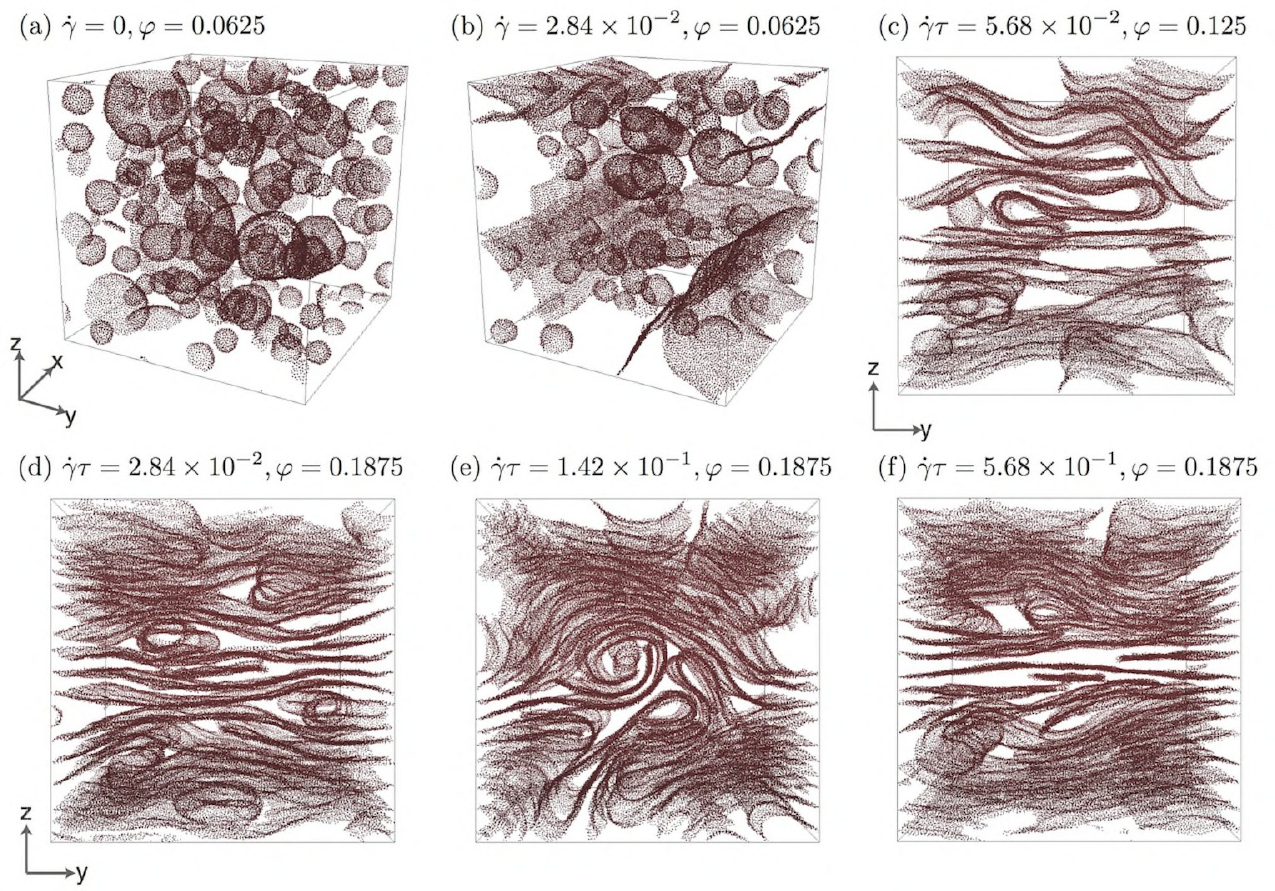}
\caption{\label{fig:M6_ves}
Snapshots of the configuration of membrane particles
for $\varphi = 0.0625$ (a) without shear ($\dot{\gamma}=0$) and 
(b) under shear flow with shear rate $\dot{\gamma}\tau=0.0284$, 
and (c) for $\varphi = 0.125$ with $\dot{\gamma}\tau = 0.0568$.
Snapshots for $\varphi=0.1875$ under shear flow 
are shown with shear rates (d) $\dot{\gamma}\tau = 0.0284$, 
(e) $\dot{\gamma}\tau = 0.142$, and (f) $\dot{\gamma}\tau = 0.568$. 
In (b)-(f), the (average) imposed flow velocity is 
${\bf v} = \dot\gamma z {\bf e}_x$. 
Solvent particles are not displayed. 
}
\end{figure*}

We now employ the meshless-membrane
model with explicit solvent to study structure formation 
in surfactant-water mixtures, both with and without shear flow.
In all simulations, the total particle number is fixed at
$N=N_\mathcal{A}+N_\mathcal{B}= 960\ 000$, 
and thus, the system size is a cubic box with side length $L=114.47\sigma$.
Simulations are performed for various membrane volume fraction
$\varphi = N_\mathcal{A}/N$, 
with $\varphi = 0.0625$, $0.125$, $0.1875$, $0.25$, and $0.3125$.
The dynamical evolution is integrated over a total time interval of
$1.2\times 10^7$ MD steps, corresponding to $2.11\times 10^3\tau$. 
All the particles of both species are initially distributed randomly
in the simulation box. 
Averages are calculated over the last $2\times 10^6$ steps, where the
system is assumed to have reached a stationary state. 
After briefly explaining the structures 
obtained by equilibrium simulations without shear in Sec.~\ref{sec:noshear}, 
we present results for the structure formation in a system under linear 
shear flow in Secs.~\ref{sec:pd} and \ref{sec:roll}.

\subsection{Mesophases in Thermal Equilibrium}
\label{sec:noshear}

\begin{figure}
\includegraphics[width=0.7\linewidth, bb = 0 0 228 418]{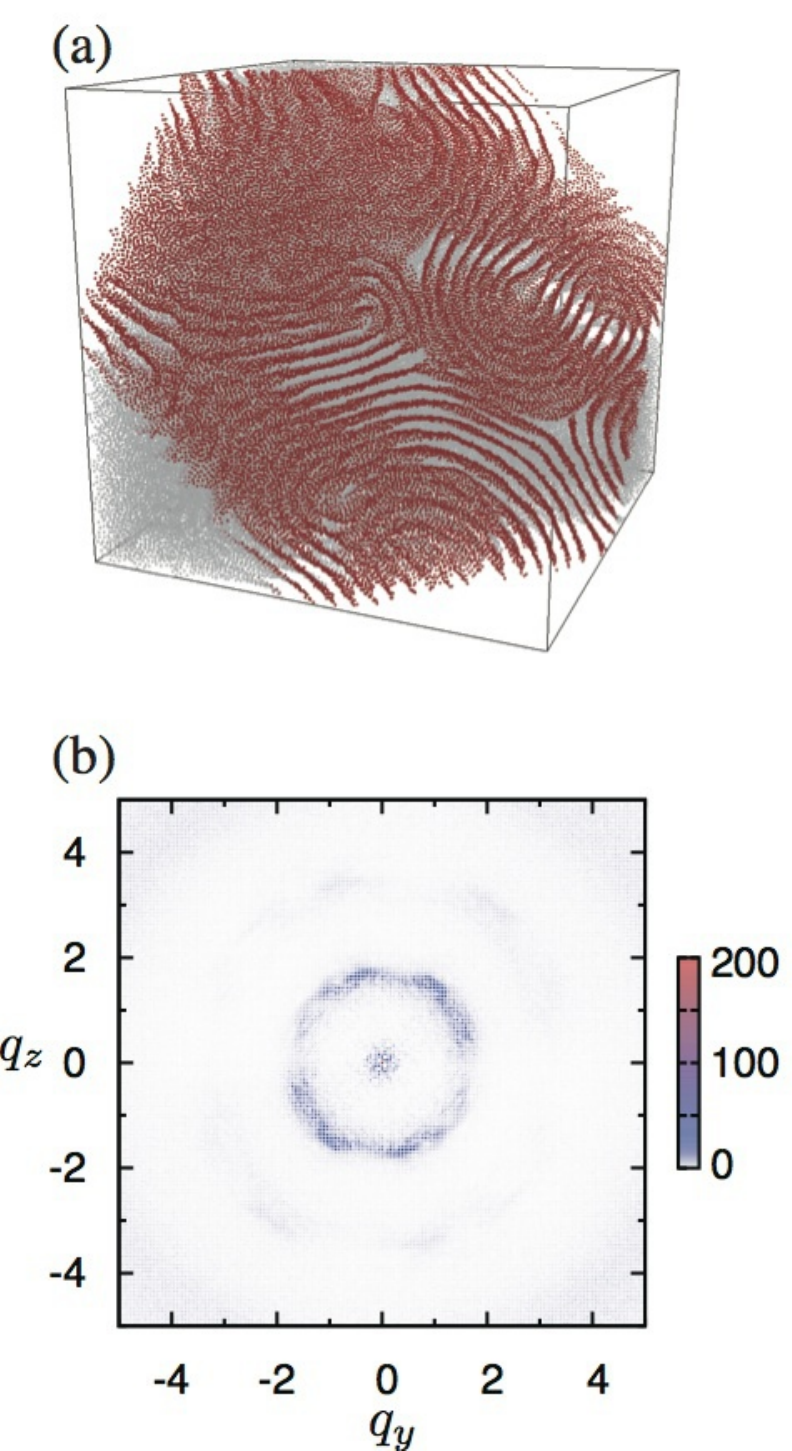}
\caption{\label{fig:noshears}
(a) Snapshot of the configuration of membrane particles 
for $\varphi = 0.3125$ without shear flow ($\dot{\gamma}=0$). 
To facilitate visualization, only a thin planar slice
is displayed. Solvent particles are not shown.  
(b) Structure factor $S_{\mathcal{A}}[\bm{q} = (0,q_y,q_z) ]$ 
for $\varphi =0.3125$ at $\dot{\gamma}=0$.   
}
\end{figure}

At a low membrane volume fraction $\varphi = 0.0625$, 
membrane particles self-assemble into vesicles,
each of which is composed of around $100$ membrane particles
(see Fig.~\ref{fig:M6_ves}(a)). 
This result is consistent with Eq.~(\ref{eq:r_c}), which 
predicts the critical particle number of a vesicle to be
$N_c = \pi R_c^2 / a^0 \simeq 75$. 
At a higher membrane volume fraction
$\varphi = 0.125\ (N_\mathcal{A}=120\ 000)$,
the membrane surfaces is found to percolate through the whole system 
via the periodic boundaries. 
This behavior is not unexpected, because each disk with radius $R_c$ 
covers a region of volume $v_c = (2R_c)^3$ by rotational diffusion,
and therefore these disks should overlap and merge 
when more than $n_c\sim V/v_c =  1200$ vesicles are present, which 
exceeds the number of vesicles with critical size ($N_\mathcal{A}/N_c$). 

For $\varphi = 0.25$ and 0.3125, periodic lamellar states
are formed owing to the repulsive interactions between the membranes,
as shown in Fig.~\ref{fig:noshears}(a). 
The membranes are curved to fill up space with random orientations. 
In thermal equilibrium, these lamellar layers 
would probably have a unique orientation throughout the system;
this well-ordered state is difficult to reach in simulations because 
the structural relaxation time well exceeds the accessible simulation 
time scale.
The 3D structure factor of the membrane density is calculated as
\begin{equation}
S_{\mathcal{A}}(\bm{q}) = \int d\bm{r}\ e^{i\bm{q}\cdot\bm{r} } 
\langle \delta \hat{n}_\mathcal{A} (\bm{r}) \delta \hat{n}_\mathcal{A} (\bm{0} ) \rangle ,
\label{eq:sk_a}
\end{equation} 
where 
$\delta \hat{n}_\mathcal{A} (\bm{r}) = 
    \sum_{i\in \mathcal{A}} \sigma^3\delta (\bm{r} - \bm{r}_i) -\phi$ 
is the local deviation of the membrane particle density from its average. 
Figure \ref{fig:noshears}(b) shows $S_\mathcal{A} (\bm{q})$
for $\varphi = 0.3125$.
The scattering intensity is spherically symmetric, as demonstrated by
a two-dimensional (2D) color-map for $q_z=0$.
Peaks arising from the inter-lamellar distance are observed 
at $|\bm{q}| = q_1= 1.74\sigma^{-1}$ and $q_2=3.49\sigma^{-1} = 2q_1$ 
with heights 9.8 and 0.78, respectively. 
The former corresponds to length of $L = 2\pi / q_1 = 3.61\sigma$,
which provides a precise estimate of the interlayer distance.

\subsection{Dynamic Phase Diagram under Shear Flow}
\label{sec:pd}

\begin{figure}
\includegraphics[width=0.9\linewidth, bb=0 0 219 195]{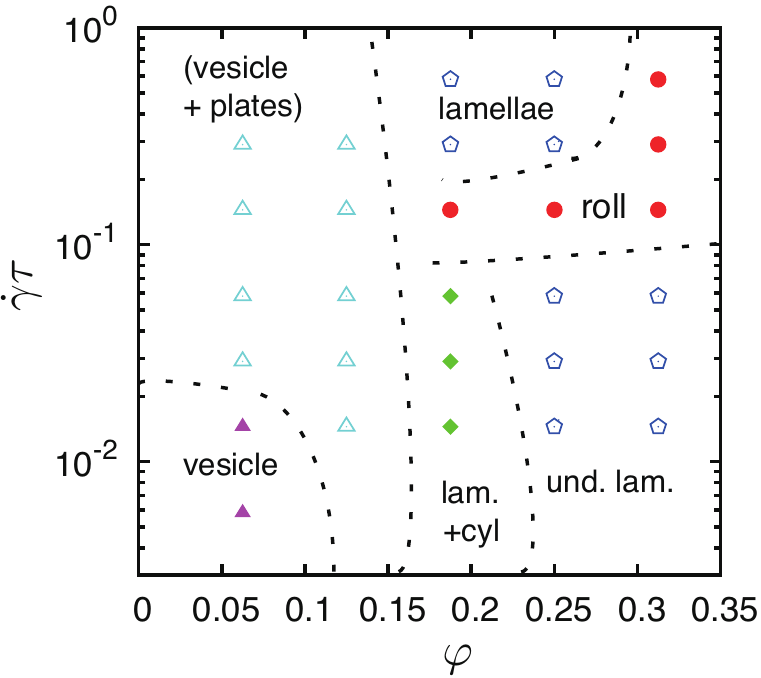}
\caption{\label{fig:phase}
Dynamic phase diagram of the explicit-solvent meshless-membrane model as 
a function of the volume fraction $\varphi$ of the membrane component
and the shear rate $\dot{\gamma}$. The parameters are
$k_\alpha =5$, $\epsilon =4$, $\phi=0.64\sigma^{-3}$, and $N=960,000$. 
The dashed lines are guides to the eye. 
}
\end{figure}

In shear flow, the vesicle and lamellar states depend 
on the concentration $\varphi$ and the shear rate $\dot{\gamma}$,
as displayed in Fig.~\ref{fig:phase}.  
At $\varphi =0.0625$, assemblies of uni-lamellar vesicles
are observed for $\dot{\gamma}\tau< 0.02$. 
At a higher shear rate, the membrane particles tend
to assemble into plate-like membrane disks, which then align parallel 
to the shear flow direction, as demonstrated by the comparison of 
snapshots in Figs.~\ref{fig:M6_ves} (a) and (b).
At $\varphi = 0.125$, lamellar layers with non-uniform 
lamellar distances are observed for $\dot{\gamma}\tau < 0.06$, as 
shown in Fig.~\ref{fig:M6_ves}(c).
There is a larger variety of phases at $\varphi = 0.1875$;
at low shear rates $\dot{\gamma}\tau < 0.1$, 
the system is a mixture of lamellae and cylinders 
as shown in Fig.~\ref{fig:M6_ves}(d);
at $\dot{\gamma}\tau = 0.142$, the lamellae 
are rolled up collectively, see Fig.~\ref{fig:M6_ves}(e); 
finally, at large shear rates $\dot{\gamma}\tau \gtrsim 0.3$, 
the system exhibits a reentrant lamellar state, 
as shown in Fig.~\ref{fig:M6_ves}(f).

\begin{figure*}
\includegraphics[width=0.75\linewidth, bb = 0 0 346 225]{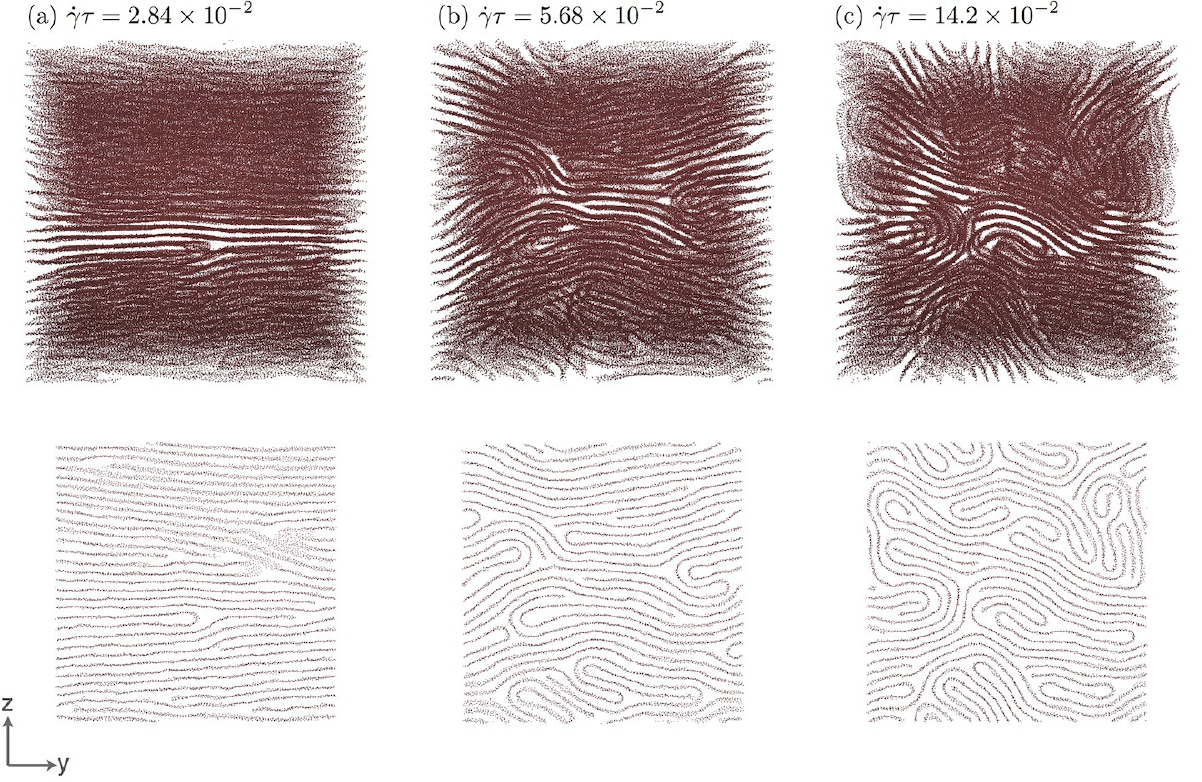}
\caption{ \label{fig:roll}
Snapshots membrane conformations for volume fraction 
$\varphi =0.3125$, with shear rates (a) $\dot{\gamma}\tau= 0.0284$,
(b) $\dot{\gamma}\tau = 0.0568$, and (c) $\dot{\gamma}\tau = 0.142$.
Views from the flow ($x$) direction are shown in the upper panels. 
Corresponding cross-sectional views (with $-3.0\sigma < x < 3.0\sigma$) 
are shown in the lower panels.  
}
\end{figure*}

At higher $\varphi$, the lamellar states align perpendicularly to 
the shear-gradient ($z$) direction in the regime of
low shear rates ($\dot{\gamma}\tau < 0.1$).
At  larger $\dot{\gamma}$, they exhibit an instability 
to a rolled-up shape whose axis is parallel to the flow direction, 
which will be investigated in more detail in Sec.~\ref{sec:roll} below. 
At $\varphi=0.25$, there is again a reentrant behavior of lamellar states, 
{\em i.e.} nearly planar aligned layers appear at large shear rates
$\dot{\gamma}\tau \ge 0.284$. 
Note that experimental phase diagrams 
often exhibit reentrant behaviors \cite{93RouxEPL,93Diat} with increase 
 in the shear rate at a certain membrane volume fraction;
the mixture changes from lamellar state to onion state,
and then after going through the coexistence region 
it enters again into an oriented lamellar state again. 
Although the onion state with densely packed MLVs
has not been obtained in our simulations, the 
reentrant behavior is qualitatively consistency 
with the experimental observations.

\subsection{Rolled-up Lamellar Structures}
\label{sec:roll}

\subsubsection{Structure Analysis}  

Snapshots of membrane conformations are shown in Fig.~\ref{fig:roll} for
$\varphi = 0.3125$ at various shear rates 
$\dot{\gamma}\tau= 0.0284$, $0.0568$, and $0.142$.
In all simulations for $\varphi \ge 0.125$, 
the membranes are completely aligned with the flow direction. 
Thus, as shown in the bottom panel of  Fig.~\ref{fig:roll},
the membrane configurations can be visualized by cross-sectional slices
perpendicular to the flow direction. 
Figure \ref{fig:roll} demonstrates the transition from the lamellar state 
to the rolled-up state, which is stable in the region $\varphi \gtrsim 0.175$ 
and $\dot\gamma\tau \gtrsim 0.1$ in the phase diagram of Fig.~\ref{fig:phase}. 

The structural changes accompanying this instability can be characterized
by the average orientation  and the mean-square local curvature of the membrane.  
The normal unit vector $\bm{n}$ is calculated from 
the first-order moving least-squares (MLS) method \cite{Nog06PRE,bely96,lanc81}
applied to the configurations of membrane particles. 
Using a weighted gyration tensor 
$a_{\alpha\beta} = \sum_j (\alpha_j' -\alpha_G' ) 
(\beta_j' -\beta_G') w_{\rm cv} (r_{ij})$, where 
$\alpha,\ \beta\in \{x, y,z\}$, 
$\bm{n}$ is obtained as an eigenvector corresponding 
to the minimum eigenvalue of $a_{\alpha\beta}$, 
which together with the other two eigenvectors $\bm{e}_1$ and $\bm{e}_2$ 
constitutes an orthonormal basis. 
Here, the cut-off function $w_{\rm cv} (r_{ij})$ in Eq.~(\ref{eq:wcv})
is employed with the same cut-off lengths $r_{cc}'=3.2\sigma$. 
As a quantitative measure for the undulation instability,
we employ the orientational order parameter  
\begin{equation}
S_z = [2({\bm{n}}\cdot \hat{\bm{e}}_z )^2-1 ]
= \cos(2\theta)   
\label{eq:angle}
\end{equation}
with the normalization 
for the two-dimensional order of cylindrical and planar symmetry.

\begin{figure}
\includegraphics[width=0.85\linewidth, bb =  0 0 341 385]{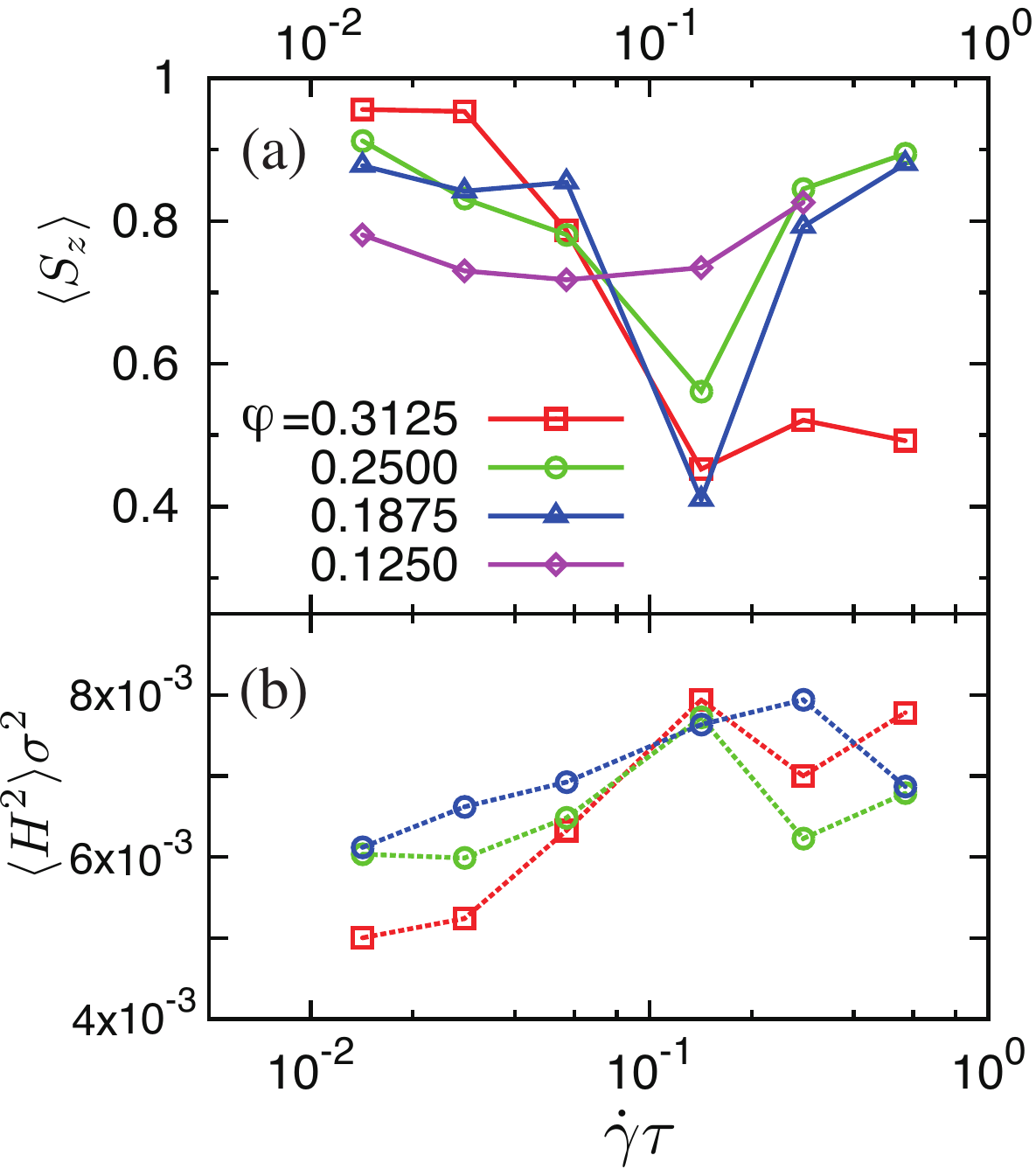}
\caption{\label{fig:angle}
(a) Membrane orientational order parameter 
$\langle S_z\rangle$ in Eq.~(\ref{eq:angle}) 
for $\varphi =0.125$, $0.1875$, $0.25$, and $0.3125$,  
and 
(b) the mean-square local curvature $\langle H^2 \rangle$ 
for $\varphi =0.1875$, $0.25$, and $0.3125$, both as 
function of the shear rate $\dot{\gamma}\tau$. 
}
\end{figure}

\begin{figure*}
\includegraphics[width=0.8\linewidth, bb = 0 0 527 205]{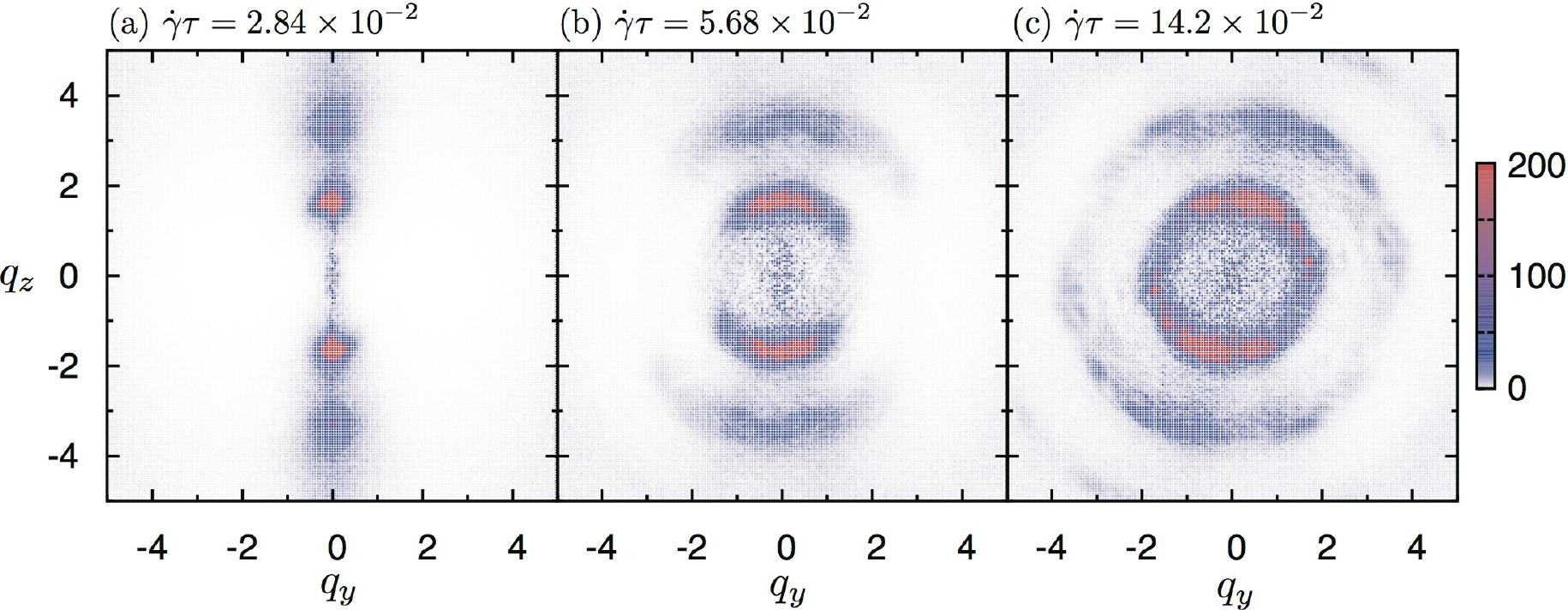}
\caption{\label{fig:str}
Color maps of the structure factor $S_\mathcal{A}(0,q_y,q_z)$ 
for volume fraction $\varphi = 0.3125$ and shear rates  
(a) $\dot{\gamma}\tau =0.0284$, (b) $\dot{\gamma}\tau =0.0568$, and 
(c) $\dot{\gamma}\tau =0.142$. 
Since the structure is uniform in the flow ($x$) direction, 
only data for $q_x =0$ are shown. 
}
\end{figure*}

The instability can also be characterized by calculating
the mean-square curvature. The second-order MLS method \cite{Nog06PRE}
provides an estimate of the membrane curvature 
from the particle configurations in the following way. 
For each particle $i$, 
 we perform a rotational transformation into
the principal coordinate system of the gyration tensor of the
neighbor particles $j$ around $i$'s weighted center 
of mass $\bm{r}_G$ by 
\begin{equation}
\left( \begin{array}{c}  
X_j \\ Y_j \\ Z_j
\end{array}\right) = \left( \begin{array}{c} 
\bm{e}_1 \\
\bm{e}_2 \\
\bm{n} \\ 
\end{array}\right)
( \bm{r}_j - \bm{r}_G)^T ,
\end{equation}
and then employ the parabolic fit function
\begin{equation}
\begin{array}{rl}
\Lambda_2(\bm{r}_i) &= \frac{1}{w_0} \sum_j \left( z_0 + h_xX_j + h_y Y_j + \frac{1}{2}h_{xx} X_j^2 \right. \\
& + \left. \frac{1}{2} h_{yy} Y_j^2  + h_{xy} X_jY_j - Z_j \right)^2 w_{\rm cv} (r_{ij}),
\end{array}
\end{equation}
where the coefficients of the Taylor expansion  $z_0,\ h_x,\ h_y,\ h_{xx},\ h_{yy}$ and $h_{xy}$ 
are fitting parameters. \cite{Nog06PRE}
By a least-squares fit, the estimated value of the mean curvature $H = (C_1+C_2)/2$ 
for particle $i$ is then obtained as 
\begin{equation}
H = \frac{(1+h_x^2) h_{yy} + (1+h_y^2 )h_{xx} - 2h_xh_yh_{xy} }{2 (1+h_x^2+h_y^2)^{3/2}}. 
\end{equation}

Figure \ref{fig:angle} displays the results for the spatial average $\langle S_z\rangle$ of the 
2D orientational order parameter  
and for the mean-square local curvature $\langle H^2 \rangle$.
When the membranes are aligned with the $x-y$ plane orthogonal to the shear-gradient direction, $S_z$ 
becomes unity.  As the membranes roll up, $S_z$ decreases.
Here, $\langle S_z\rangle =0$ for perfectly 
cylindrical state (where  $({\bm{n}}\cdot \hat{\bm{e}}_x)^2  =0$ and
$({\bm{n}}\cdot \hat{\bm{e}}_y)^2 =({\bm{n}}\cdot \hat{\bm{e}}_z)^2 =1/2$ ), and $\langle S_z\rangle = -1$ 
for a perfectly flat lamellar layers 
perpendicular to the vorticity ($y$) direction  (where $({\bm{n}}\cdot \hat{\bm{e}}_z)^2 =0$). 
For all $\varphi\ge 0.1875$, the values of $\langle S_z\rangle$ and $\langle H^2\rangle$
provide evidence for rolled-up structures at $\dot{\gamma}\tau = 0.142$. 
For $\varphi = 0.1875$ and 0.25,
$\langle S_z\rangle$ increases 
again at $\dot{\gamma}\tau = 0.284$, indicating reentrant 
alignment of the membranes in a lamellar stack. 
However, for $\varphi = 0.3125$, $\langle S_z\rangle$ remains small
(and $\langle H^2 \rangle$ large),
which implies that rolled-up structures exist also at 
$\dot{\gamma}\tau = 0.284$. 
Thus, the evolution of rolled-up conformations is 
the most pronounced at the highest membrane volume fraction $\varphi =0.3125$.

In Fig.~\ref{fig:str}, results for the structure factor $S_{\mathcal{A}}(\bm{q})$ at $q_x=0$ 
are shown for the same set of data as in Fig.~\ref{fig:roll}. Due to the nearly complete 
alignment of membranes  in the flow ($x$) direction, all the structural features are 
reflected in $S_{\mathcal{A}}(\bm{q})$ in the $q_y-q_z$ wave-vector plane.
Since lamellar layers are nearly planar  at the low shear rate $\dot{\gamma}\tau = 0.0284$, 
a sharp peak is observed around $(q_y,q_z) = (2\pi q_1,0)$. 
Because of the undulation instability, 
the pattern becomes more circular at higher $\dot{\gamma}$. 

In the small-angle neutron \cite{03Nettesheim} and X-ray \cite{11Ito} scattering experiments,
the scattering beams can be injected from
two directions, either radial or tangential to the shear cell,
which correspond to shear-gradient and flow directions, respectively.
After the shear flow is applied, 
after a while a  Bragg peak of the radial beam develops in the vorticity direction, 
while the scattering pattern becomes isotropic for the tangential beam.
This suggests  2D-isotropic undulations of the
lamellar structure perpendicular to the flow direction.
Later in time, the radial beam is scattered isotropically 
in the tangential direction, indicating the formation of onion structures. 
The  structure factor $S(\bm{q})$ in our simulation (Fig.~\ref{fig:str}(c)),
agrees well with $S(\bm{q})$ of this transient states in these 
scatting experiments.
Measurements of solvent diffusion in a Rheo-NMR experiment of a non-ionic 
surfactant system \cite{08Medronho} show a 
diffusion anisotropy  in the direction of shear flow in the intermediate state. 
These experimental results suggest 
that the membranes are aligned in the direction of shear 
flow in the intermediate state of the lamellae-to-onion transition,
although more detailed information on the structural membrane arrangement could
not be obtained. 
The rolled-up lamellar structures observed in our simulations
match the experimental evidence, so that they are a good
candidate for the intermediate states.

\subsubsection{Temporal Evolution of Membrane Structures}
\label{sec:ev}

In simulations of a large system, even if extended over a 
long time interval, the resultant structure often exhibits dependence on the initial 
conditions -- owing to the slow processes involved in the dynamics. 
Therefore, we compare here the time evolution at $\varphi =0.3125$
starting from both, a lamellar state and a random distribution of membrane
particles (the latter corresponding to the simulations described in the previous subsections). 
For an initial lamellar state, the final configuration of a simulation run with 
$1.2\times 10^7$ MD steps at a small shear rate $\dot{\gamma}\tau = 0.0284$ (see Fig.~\ref{fig:roll}(a)), is employed.

\begin{figure}
\includegraphics[width=0.80\linewidth, bb = 0 0 322 479]{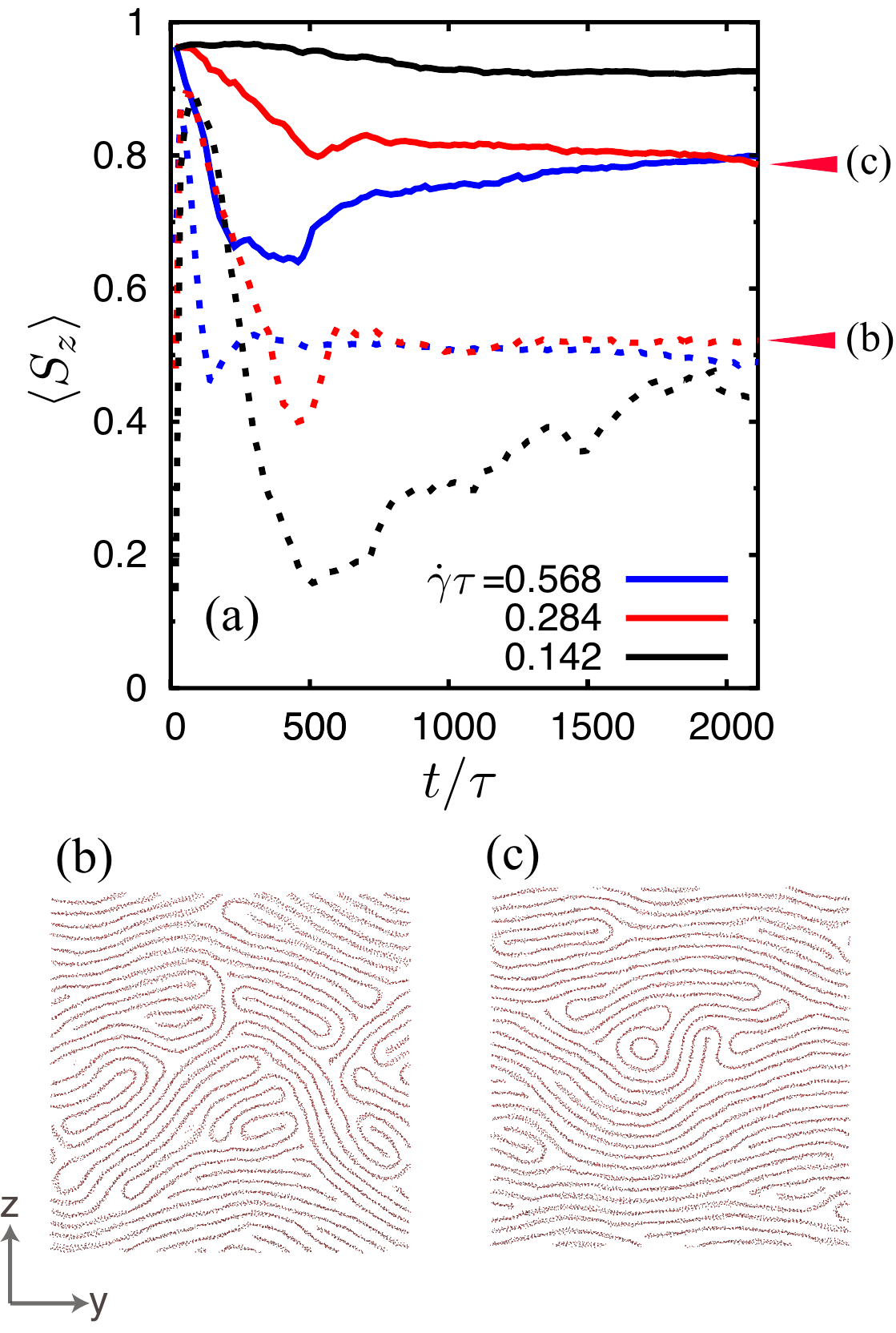}
\caption{\label{fig:tdep}
Comparison of structure formation from a random initial configuration
and from  a lamellar state at $\varphi =0.3125$.
(a) Time evolution of the orientational order parameter $\langle S_z\rangle$. 
The dotted and solid lines show the data starting
from the random and lamellar initial states, respectively.
Blue, red, and black lines represent
the shear rates $\dot{\gamma}\tau = 0.568$, $0.284$, and $0.142$, 
respectively. 
Snapshots of the final configurations for $\dot{\gamma}\tau = 0.284$
at $t=2.11\times 10^3 \tau$,  
as they have developed from (b) the random and 
(c) the lamellar initial states, are also shown. 
}
\end{figure}

In Fig.~\ref{fig:tdep}(a),
the orientation order parameter $\langle S_z\rangle$
for both initial conditions are shown for 
$\dot{\gamma}\tau= 0.568$, $0.284$, and $0.142$. 
For the case of a random initial distribution,
small disks merge into 
randomly oriented surfaces, which then align in the shear flow
to become a nearly perfect lamellar stack with some defects.  
Afterwards, lamellae roll up into slightly larger rolls. 
During rolling up, $\langle S_z\rangle$ exhibits an overshoot
(see Fig.~\ref{fig:tdep}(c)), and finally approaches a constant as the
structure relaxes into a (meta)stable state.
The overshoot amplitude depends on initial states 
and random noises.   

Figs.~\ref{fig:tdep}(b) and (c) show the membrane
conformations after an elapsed time of $t=2.11\times 10^3\tau$ 
(for $\dot{\gamma}\tau = 0.284$) for the two types of initial 
conditions, and explain the origin of the substantially different
values of the order parameter $\langle S_z\rangle$ in 
Fig.~\ref{fig:tdep}(a). When the random state is taken as initial 
condition, rolled-up structures are considerably more pronounced;
this may be traced back to the presence of defects in the lamellar
structure which forms at short times $t/\tau \simeq 100$. 

For the case of a lamellar initial configuration, 
the undulation instability becomes more 
conspicuous when the applied shear flow is stronger.
Moreover, while strong undulations are observed at 
low $\dot{\gamma}$ with the random
initial configuration, less undulations take place 
with the lamellar initial configuration, as indicated 
by $\langle S_z\rangle \sim 1$. 
Thus, the initial conditions play an important role
in the selection of the transition 
path and structure formation.
It may depend not only on the shear rate 
and relaxation time but also on the 
distribution of structural defects in the lamellar layers.
Thus, more systematic studies are required in the future 
to clarify the hysteresis of these systems.

\section{Summary}
\label{sec:sum}

In this paper, we have constructed an explicit-solvent meshless-membrane model
for surfactant-water mixtures. 
The model reproduces properties of an earlier implicit-solvent 
meshless membrane model, where membrane  
bending rigidity and line tension can be independently controlled to a large extent. 
The model enables large-scale simulations of structural changes, where dynamical effects 
of hydrodynamic interactions have to be taken into account. 
At present, such a large simulation with 
as many as one-million particles 
can be realized by parallelized molecular dynamics simulation methods. 

Our main results concern the effects of shear flow on the structure formation
of membrane ensembles. 
Various structures including vesicles, lamellae, and multi-lamellar states with 
nearly cylindrical symmetry have been found, most of which are qualitatively
consistent with experimental observations of non-ionic surfactant membranes
under shear flow. 
Especially, a cylindrical instability of multi-lamellar 
membrane is predicted to occur perpendicularly to the flow direction.
The corresponding scattering patterns are in qualitatively agreement with the 
results of small angle neutron (and X-ray) scattering experiments 
under shear deformation. 
The rolled-up lamellae are a good candidate for the intermediate structures 
on the way to the onion state, which are observed in the experiments. 

Our simulations do not reproduce onion formation, which is ubiquitously observed
in experiments on $\mu$m length scales.
We speculate that it is due to the limited system size, which can be 
overcome by larger-scale calculations in the future.  On the other hand, in the experiments, 
strains larger than $\dot{\gamma}t \gtrsim 10^4$ are necessary to reach the onion state, 
which indicates that very long simulation runs are required to obtain these states. 

The control of the physical parameters including the line tension $\Gamma$, the 
bending rigidity $\kappa$, and the Gaussian modulus $\bar{\kappa}$
is another challenge. 
While $\Gamma$ and $\kappa$ are easily controlled in our model, $\bar{\kappa}$ is more difficult.
The Gaussian modulus $\bar{\kappa}$ might also
play an important role in the structure formation, because it is directly related to topological
changes happening on the way of onion formation. \cite{2012Hu}

\begin{appendix}
\section{Solvent-Mediated Forces at Higher Solvent Density}

In the present simulation model,
solvent particles have a similar size as membrane particles.
Here, we discuss the finite-size effects of the solvent particles,
if much higher solvent densities are used than in the present study.
When the solvent particles are densely packed, the system approaches a crystallization
transition, and local crystalline order emerges to bring about
interactions between closely spaced lamellar layers. 

As an example, we study the explicit-solvent meshless-membrane model for higher 
number density  (denoted system $(\mathcal{S}2$), where
$\phi = (\sigma_{\mathcal{A}}^3N_{\mathcal{A}} 
+ \sigma_{\mathcal{B}}^3N_\mathcal{B}) /V = 0.72$ 
with the total particle number $N=N_{\mathcal{A}} + N_{\mathcal{B}}=
480\ 000$). The particle radii of the two components 
are chosen such as $\sigma_{\mathcal{B}}
=1.2\sigma_{\mathcal{A}}$, where $\sigma_{\mathcal{A}}$ 
and $\sigma_{\mathcal{B}}$ denote the radii of
membrane  and solvent components, respectively. 
The cut-off lengths of the interactions 
are set to $r_c^{\rm rep} =2.7\sigma_{\mathcal{A}} , r_{\rm ga}=1.5\sigma_{\mathcal{A}},\ 
r_{\rm cc}=3.0\sigma_{\mathcal{A}}$, respectively.  
The repulsive inverse twelfth-power potential exhibits melting 
transition around the volume fraction around 0.43 (corresponding to $\phi \simeq 0.8$
in our definition). \cite{70Hoover}
Thus, our density $\phi =0.72$ is close to the crystallization line.

\begin{figure}
\includegraphics[width=0.65\linewidth, bb=0 0 474 460]{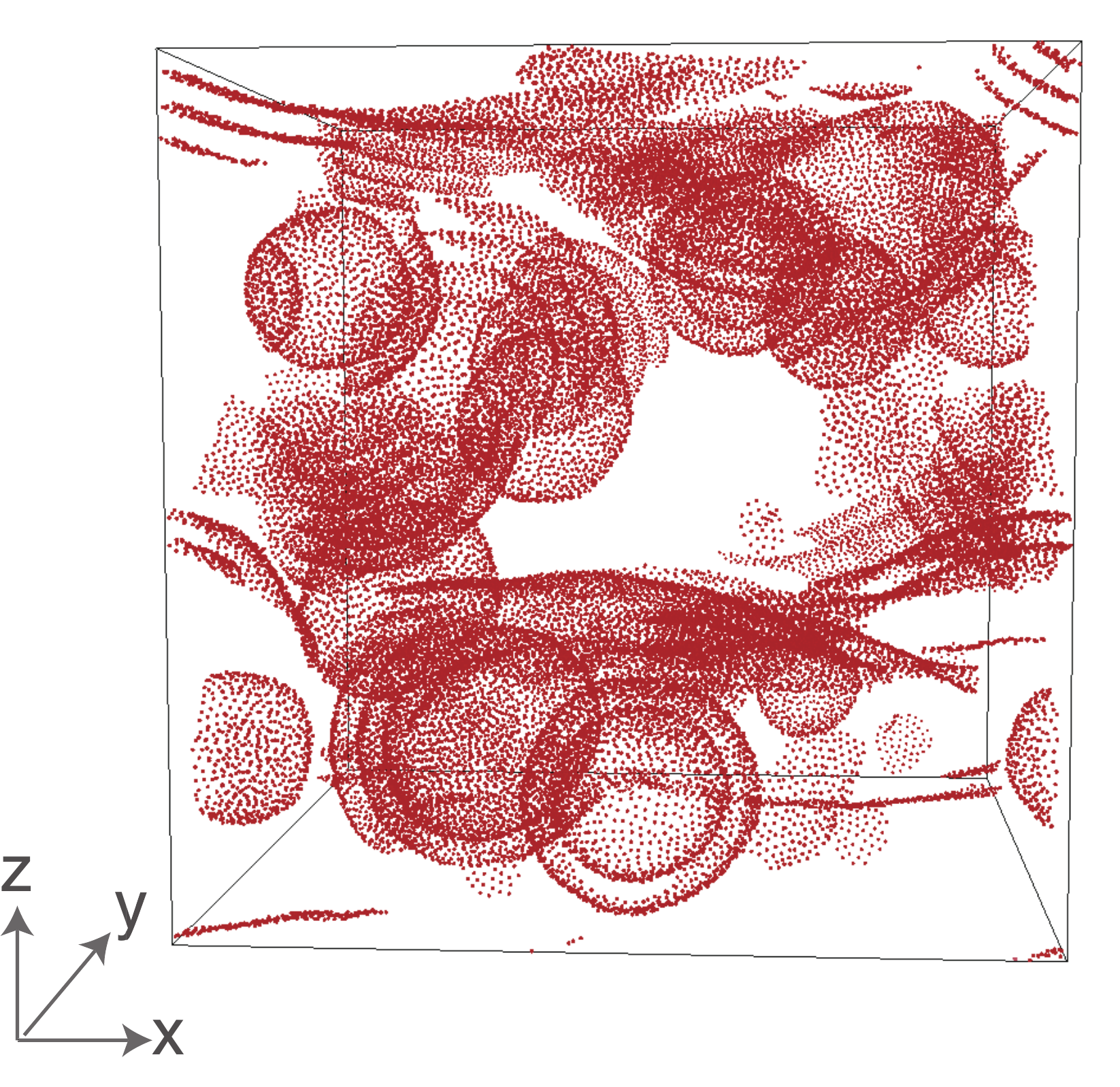}
\caption{\label{fig:mlv_dep}
Snapshot of the MLV state due to 
the solvent-mediated attractive interactions for the system $\mathcal{S}2$
at $\phi = 0.72$,
$\varphi = 0.06$ and $N=480,000$ with different 
size ratio $\sigma_{\mathcal{B}}/\sigma_{\mathcal{A}} = 1.2$, 
for shear rate $\dot{\gamma}\tau = 0.0142$. 
}
\end{figure}

As an example, Fig.~\ref{fig:mlv_dep} shows a snapshot
for membrane volume fraction 
$\varphi = N_\mathcal{A}\sigma_{\mathcal{A}}^3 / (N_{\mathcal{A}}\sigma_{\mathcal{A}}^3 
+ N_{\mathcal{B}}\sigma_{\mathcal{B}}^3) = 0.06$
for a shear rate $\dot{\gamma}\tau = 0.0142$.
After an initial relaxation from a random configuration of the 
membrane particles, the membrane layers assemble into
stacks, and then form MLVs after gathering a certain amount of the membrane patches.  
In the snapshot, the layers are stacked with distances $d\simeq 3.0\sigma_{\mathcal{B}}$,
which seems to arise from the discreteness of the interstitial solvent particles
-- here it should be noted that interlayer distance cannot be smaller than 
$d \simeq 2.0\sigma_{\mathcal{B}}$
because it is inside the range of $U_{\rm rep}$ and $U_\alpha$, which both act repulsively.

\begin{figure}
\includegraphics[width=0.65\linewidth, bb=0 0 227 222]{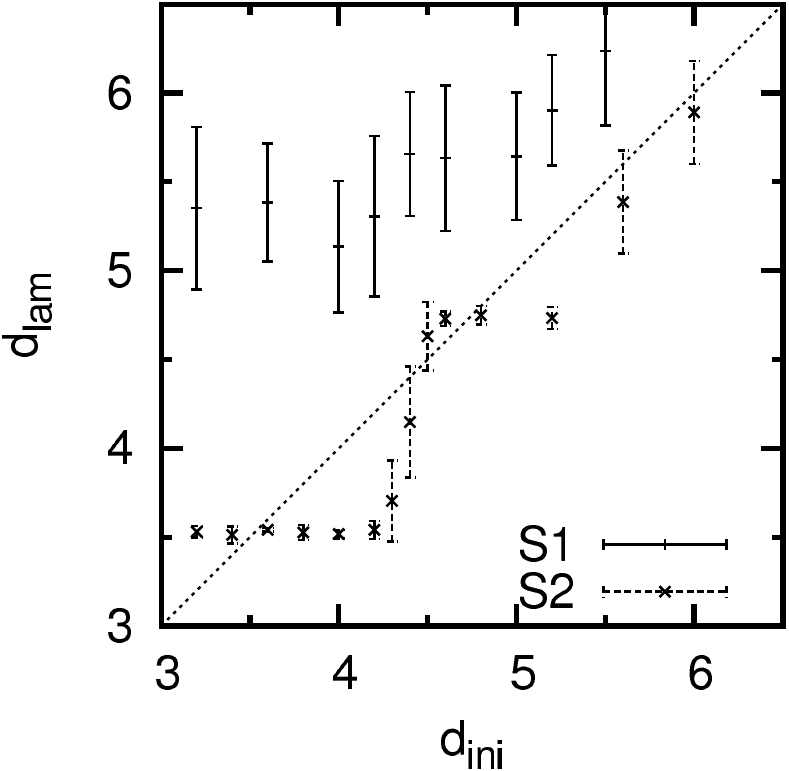}
\caption{\label{fig:SLdist}
Equilibrium lamellar distance $d_{\rm lam}$
versus initial lamellar distance $d_{\rm ini}$ 
between a tension-less planar membrane and two (smaller) membrane discs.
For the systems $\mathcal{S}1$ and $\mathcal{S}2$, 
lengths are represented in units of $\sigma$ and $\sigma_{\mathcal{A}}$,
respectively. 
The dotted line indicates the line $d_{\rm ini}=d_{\rm lam}$. 
}
\end{figure}

To avoid this solvent-mediated force, a longer cut-off range of the 
repulsive forces and a lower density are employed
in the simulation described in the main text (denoted system ($\mathcal{S}1$)). 
To demonstrate the difference between the systems $\mathcal{S}1$ and $\mathcal{S}2$, 
we simulate three lamellar layers in the following way. 
A periodic simulation box with lengths 
$L_x=L_y  = \sqrt{A_{xy}^0}/\sigma_{\mathcal{A}}$,
$L_z  = V/ (L_xL_y)$
is set up so that a tension-less membrane (with particle
number $N_{\mathcal{A}0}) = 1600$ is put along the 
$xy$-plane. Here, the projected areas
are set to $A_{xy}^0 = 1.267N_{\mathcal{A}0}$ 
for $\mathcal{S}1$ and $A_{xy}^0 = 1.335N_{\mathcal{A}0}$ 
for $\mathcal{S}2$, respectively. 
With various interlayer distance $d_{\rm ini}$, circular membrane disks 
with particle number $N_{\mathcal{A}1}=N_{\mathcal{A}2}=400$
are put on both sides of the membrane. 
Solvent particles are then inserted to fill up the system; they
are relaxed for fixed membrane configuration for $10^5$ MD steps. Afterward,  
a simulation of the full system is performed for $1.5\times 10^6$ MD steps 
to obtain an equilibrium interlayer distance $d_{\rm lam}$. As shown in Fig.~\ref{fig:SLdist}, 
while the interlayer distance increases with time 
in the $\mathcal{S}1$ case, there are stable lamellar distances 
around $d_{\rm lam} = 3.5\sigma_{\mathcal{A}}$ and $4.7\sigma_{\mathcal{A}}$
in the $\mathcal{S}2$ case with higher solvent density, 
which correspond to $3\sigma_{\mathcal{B}}$ and $4\sigma_{\mathcal{B}}$, respectively.
Thus, in $\mathcal{S}2$, the effective attractive potentials
lead to the stable multi-lamellar layers due to the attractive interactions.

\section{Numerics and Parallelization}

\begin{figure}
\includegraphics[width=0.9\linewidth, bb = 0 0 563 496]{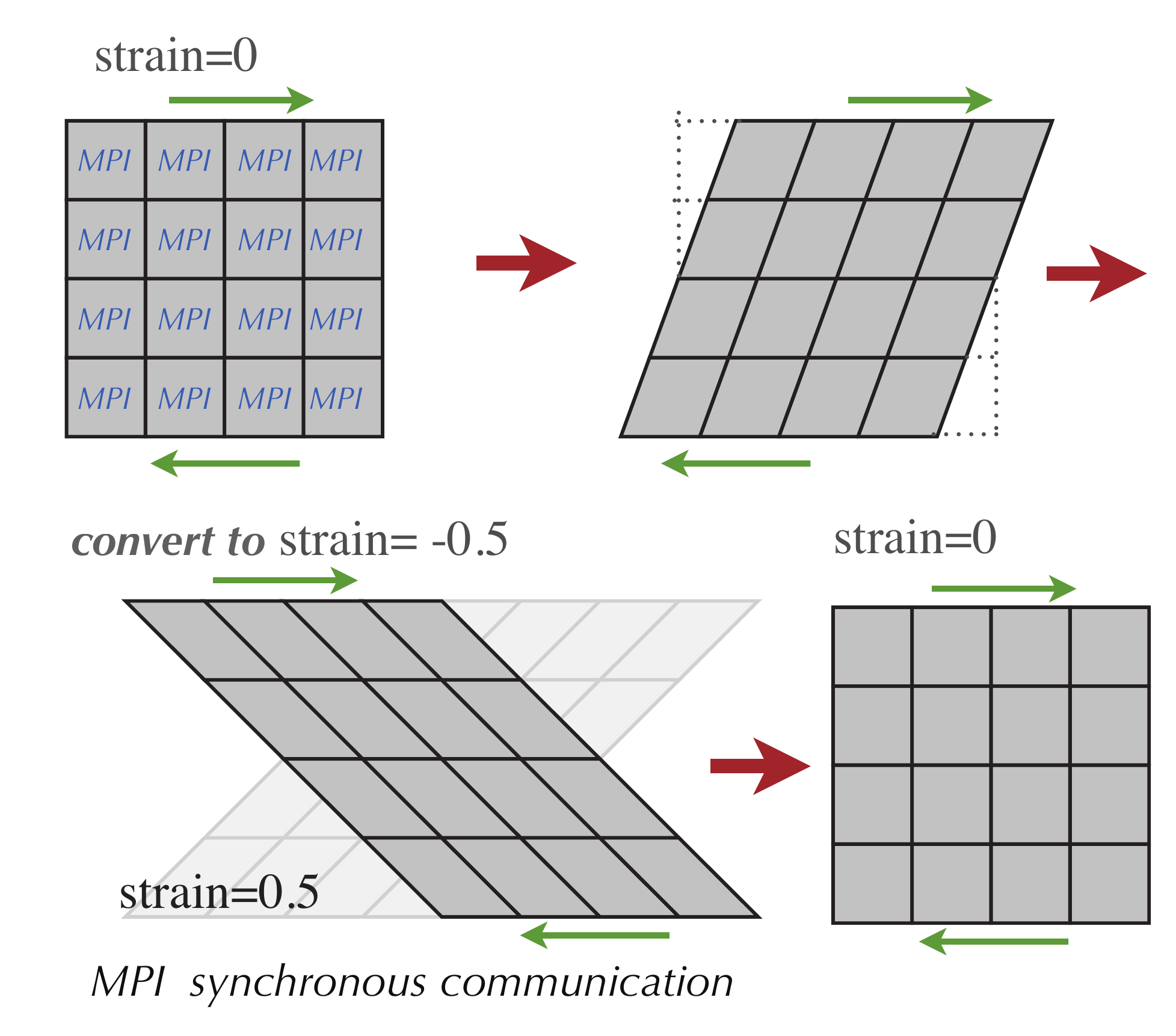}
\caption{\label{fig:LE}
Schematic picture in two dimensions 
for the implementation of Lees-Edwards boundary conditions in 
a program parallelized with MPI communication. 
}
\end{figure}

Numerical simulations have been carried out on massively parallel supercomputers.
On 256 CPUs of Intel Xeon X5570 (2.93GHz) in SGI Altix ICE 8400EX at ISSP, 
it costs 72 hours to perform a run of $1.2\times 10^7$ simulation steps
for $\varphi = 0.3125$ and $N=960,000$. 
Here, about 10\% of the total time is for the force calculation 
of $U_{\rm rep}$, 30\% for $U_\alpha$, around 20\% for the construction of the buffer, 
and 20\% for the communication between the MPI processes. 
The system is divided into cubic (or rectangular) boxes, 
each of which is calculated by one MPI process.
We also parallelize  calculations of each process with
the use of OpenMP by performing calculations of different far away pairs 
at once on different threads.
The code is optimized to achieve 
15\% performance compared with the theoretical limit
on X5570 processors. 

Each process has separate cell lists and neighbor 
lists for solvent and membrane particles. 
To apply shear, we employ a simultaneous affine deformation of 
the total system, each MPI process box,
and each cell for neighbor search, so that a square is transformed   
into a parallelogram shape consistent with the shear deformation, as
illustrated in Fig.~\ref{fig:LE}. When the strain 
reaches 0.5, these parallelograms are reflected 
to make the strain -0.5, see Fig.~\ref{fig:LE}, which basically requires 
an all-to-all communication of all the position 
and velocity data.

\end{appendix}

\begin{acknowledgments}
This work was supported by Grant-in-Aid for Young Scientists 24740285 from JSPS in Japan, 
Computational Materials Science Initiative (CMSI) from MEXT in Japan, 
and the European Soft Matter Infrastructure project ESMI (Grant No. 262348) in the EU. 
We would like to thank S. Fujii, S. Komura, H. Watanabe,
U. Schiller, M. Peltom\"aki, G.A. Vliegenthart, 
and D. Y. Lu for informative discussions. 
The numerical calculations were carried out on 
SGI Altix ICE 8400EX and NEC SX-9 at ISSP in University of Tokyo (Japan),
Fujitsu FX10 at Information Technology Center in University of Tokyo (Japan),
Hitachi SR16000 at YITP in Kyoto University (Japan),
and JUROPA at J\"ulich Supercomputing Center at Forschungszentrum J\"ulich (Germany). 
\end{acknowledgments}

\vskip5mm
\hspace{-0.3cm}


\end{document}